\documentclass{JHEP3}
\usepackage{amsmath}
\newcommand{\ep}{\varepsilon}

\newcommand{\Li}[2]{{\mbox{Li}}_{#1}\left(#2\right)}

\newcommand{\Snp}[2]{{\mbox{S}}_{#1\!}\left(#2\right)}
\title{
On the all-order epsilon-expansion of generalized hypergeometric functions
with integer values of parameters
}
\author{
Mikhail~Y.~Kalmykov\\
Present address: II. Institut f\"r 
Theoretische Physik, Universit\"at Hamburg\\
Luruper Chausee 149, 22761
Hamburg, Germany\\
E-mail: \email{kalmykov@theor.jinr.ru}
}
\author{
Bennie F.L. Ward\\
Department of Physics, Baylor University, \\
One Bear Place, Box 97316, Waco, TX 76798-7316 \\
E-mail: {\tt\href{mailto:BFL_Ward@baylor.edu}{BFL\underline{\ }Ward@baylor.edu}}

}
\author{Scott A. Yost\\
Department of Physics, Princeton University,\\
Princeton, NJ 08544, U.S.A.\\
E-mail: \email{syost@princeton.edu}
}
\abstract{ \\
We continue our study of the construction of analytical 
coefficients of the epsilon-expansion 
of hypergeometric functions and their connection with Feynman diagrams. 
In this paper, we apply the approach of obtaining iterated solutions to the 
differential equations associated with
hypergeometric functions to prove the following result: \\

{\bf Theorem 1:}\\
The epsilon-expansion of a generalized hypergeometric function with 
integer values of parameters,
\[
{}_pF_{p-1}(I_1+a_1\ep, \cdots, I_{p} + a_p \ep; 
             I_{p+1}+b_1\ep, \cdots, I_{2p-1} + b_{p-1};z) \;,
\]
is expressible in terms of generalized polylogarithms with coefficients
that are ratios of polynomials. \\

The method used in this proof provides an efficient algorithm for 
calculating of the higher-order coefficients of Laurent expansion.
}
\begin{document}
%
\section{Introduction}
Hypergeometric functions are useful in the evaluation of Feynman diagrams.
See, for example, Ref.\ \cite{feynman} for a review of how these functions
arise.  In this paper, we will be concerned with
the manipulation of hypergeometric functions \cite{bateman,slater}, by which 
we understand specifically 
\begin{itemize}
\item[] (1) the reduction of the original function to a minimal set of 
basis functions,
\item[] (2) the construction of the all-order $\ep$-expansion of 
basis functions.
\end{itemize}
The $\ep$-expansion refers to the Laurent expansion of hypergeometric functions 
about rational values of their parameters in terms of known functions or 
perhaps new types of functions. In the latter case, the problem remains to 
identify the full set of functions which must be invented to construct this 
expansion for general values of the parameters.\footnote{All these procedures 
coincide with standard techniques used in the analytical 
calculation of Feynman diagrams.\,\cite{ibp,Kotikov} It has long been
expected that all Feynman diagrams can be represented by some class of 
hypergeometric functions.  Now we can propose specifically that any Feynman 
diagrams can be associated with the Gelfand-Karpanov-Zelevinskii (GKZ or A-
hypergeometric function) hypergeometric functions\,\cite{GKZ}.
Let us recall that Lauricella's, Horns' and generalized hypergeometric 
functions occur as special cases of the GKZ-systems. For an introduction, we 
recommended Ref.\ \cite{cattani}.}

Problem (1) is a purely mathematical one.  It is closely related with 
the existence of algebraic relations between a few hypergeometric functions 
with values of parameters differing by an integer, the so-called ``contiguous 
relations''\,\cite{Rainville}.  The systematic procedure for solving the
relevant recursion relation is based on the Gr\"obner basis technique.  
In particular, a proper solution for generalized hypergeometric functions, 
the so-called ``differential reduction algorithm,'' was developed by 
Takayama\,\cite{takayama}.
(See Ref.\ \cite{review} for a review.)  By a differential reduction 
algorithm, we will understand a relation of the type 
$F(\alpha \pm j,\vec{b};z)=\Pi_{k=1}^m D(\alpha+k;\vec{b}) 
F(\alpha,\vec{b};z)$, 
where $j,k$  are integers, $\vec{b}$ is a list of additional parameters and 
$D$ is a differential operator of the form 
$D=A(\alpha;\vec{b};z)\frac{d}{dz} + B(\alpha;\vec{b};z)$.\footnote{An 
algorithm of differential reduction of generalized hypergeometric functions 
to a minimal set allows the calculation of any Feynman diagram that is 
expressible in terms of hypergeometric functions without any reference to 
integration by parts or the differential equation technique. 
The application of this algorithm to the calculation of Feynman diagrams will 
be presented in another publication.} For Gauss hypergeometric functions, 
the reduction algorithm was presented in Refs.\ \cite{2F1,MKL06}.
For generalized hypergeometric function, is it equivalent to the statement 
that any function ${}_pF_{p-1}(\vec{a};\vec{b};z)$ can be expressed as 
a linear combination of functions with arguments that differ from the 
original ones by an integer, ${}_pF_{p-1}(\vec{a}+\vec{m};\vec{b}+\vec{k};z)$, 
and the function's first $p-1$ derivatives:
\begin{eqnarray}
 \hspace{-5mm}
R_{p+1}(\vec{a},\vec{b},z)&&
{}_{p}F_{p-1}(\vec{a}+\vec{m};\vec{b}+\vec{k}; z)
 = 
\nonumber \\ && \hspace{-20mm}
\Biggl \{
  R_1(\vec{a},\vec{b},z)\,\left( \frac{d}{dz}\right)^{\,p-1}
+ \cdots 
+ R_{p-1} (\vec{a},\vec{b},z)\,\frac{d}{dz}
+ R_{p}(\vec{a},\vec{b},z) 
\Biggr\}
{}_{p}F_{p-1}(\vec{a};\vec{b}; z) \;,
\label{reduction}
\end{eqnarray}
where $\vec{m}, \vec{k}$ are lists of integers, 
the $R_i$ are polynomials in parameters $a_i,b_j$ and $z$.

Problem (2) arises in physics in the context of the analytical 
calculation of Feynman diagrams.  The complete solution of this problem is 
still open.  We will mention here some results in this direction derived 
by physicists.  Let us recall that there are three different ways to describe 
special functions: 
\begin{itemize}
\item[] (i)  as an integral of the Euler or Mellin-Barnes type, 
\item[] (ii) by a series whose coefficients satisfy certain recurrence 
relations,
\item[] (iii) as a solution of a system of differential and difference equations
(holonomic approach). 
\end{itemize}
For functions of a single variable, all of these representations are 
equivalent, but some properties of the function may be more evident
in one representation than another.  These three different representations 
have led physicists to three different approaches to developing the 
$\ep$-expansion of hypergeometric functions. 

The Euler integral representation (i) was developed intensively
by Davydychev, Tarasov  and their collaborators \cite{integral-representation}
and the most impressive result was the construction of the all-order 
$\ep$-expansion of Gauss hypergeometric functions in terms of Nielsen 
polylogarithms \cite{hyper:partial}.  This type of Gauss hypergeometric 
function is related to one-loop propagator-type diagrams with arbitrary 
masses and momenta, two-loop bubble diagrams with arbitrary masses, and 
one-loop massless vertex-type diagrams. 

The series representation (ii) is a very popular and intensively studied
approach. The first results of this type were derived by David 
Broadhurst\,\cite{Broadhurst:1996} for the so-called ``single scale'' diagrams,
which are associated with on-shell calculations in QED or QCD, with further 
developments appearing subsequently in 
Ref.\ \cite{series-representation}.\footnote{Some relations between the 
Mellin-Barnes representations and series representations of Feynman
diagrams follow from the Smirnov-Tausk approach \cite{Smirnov-Tausk}.}
Particularly impressive results involving series representations were 
derived recently by Moch, Uwer, and Weinzierl in the framework of the 
nested sum approach.\cite{nested1,nested2} Algorithms based on this
approach have been implemented in computer code.\cite{code1}
The series approach leads to
algebraic relations between the analytic coefficients of the $\ep$-expansion, 
but does not provide a way to obtain a reduction of the original 
hypergeometric function before expansion.  Another limitation of this 
approach is that the parameters are restricted to 
integer values or special combinations of half-integer values 
(so-called ``zero-balance'' parameter sets).\footnote{For some new results on 
$\ep$-expansions of hypergeometric functions with nonzero-balance 
parameter sets of parameters (specifically, one half-integer parameter), 
see Ref.\ \cite{KWY07:2}.}

For approach (iii), obtaining iterated solutions to the proper 
differential equations associated with hypergeometric functions, the first 
results were obtained for Gauss hypergeometric functions expanded about
integer values of parameters.\cite{hyper:expansion}  In Ref.\ \cite{KWY07:1}, 
that result was extended to combinations of integer and half-integer values of 
parameters.  An advantage of the iterated solution approach over the series 
approach is that it provides a more efficient way to calculate each order of 
the $\ep$-expansion, since it relates each new 
term to the previously derived terms rather than having to work with an 
increasingly large collection of independent sums at each subsequent order.

The aim of the present paper is to apply approach (iii) to proving the 
following theorem:\footnote{In fact, the result expressed in this 
theorem can be proved within the nested sum approach\,\cite{nested1}. However, 
our idea is to extend the iterated solution approach to this more complicated 
system and in the process, derive a more efficient algorithm for calculating 
the analytical coefficients of the $\ep-$expansion.}\\[1em]
\noindent{\bf Theorem 1} \\
{\it 
The all-order $\ep$-expansion of a generalized hypergeometric function
${}_pF_{p-1}(\vec{A}+\vec{a}\ep;\vec{B}+\vec{b}\ep;z)$, 
where $\vec{A}$ and $\vec{B}$ are lists of integers,
are expressible in terms of generalized polylogarithms 
{$($see Eq.\ \ref{GP}$)$} with coefficients that are ratios of polynomials.
}\\[1ex]

\noindent To be specific, this means that
\[
P(\{a\},\{b\},z)\ {}_pF_{p-1}(\vec{A}+\vec{a}\ep;\vec{B}+\vec{b}\ep;z) = 
\sum R_{\vec{s}}(z)\; \Li{\vec{s}}{z} \ep^k\;,
\]
where $\vec{s} = (s_1, \cdots, s_l)$ is a multiple index
and $P(\{a\},\{b\},z), R_{\vec{s}}(z)$ are polynomials.
The {\it generalized polylogarithms} are defined by the equation
\begin{equation}
\Li{k_1,k_2, \cdots, k_n}{z} = 
\sum_{m_1 > m_2 > \cdots m_n > 0} \frac{z^{m_1}}{m_1^{k_1} m_2^{k_2} \cdots m_n^{k_n}} \;,
\label{GP}
\end{equation}
For completeness, we 
recall that {\it generalized polylogarithms} (\ref{GP}) can be expressed as 
iterated integrals of the form
\begin{eqnarray}
\Li{k_1, \cdots, k_n}{z} & = &  
\int_0^z 
\underbrace{\frac{dt}{t} \circ \frac{dt}{t} \circ \cdots \circ 
\frac{dt}{t}}_{k_1-1 \mbox{ times}} \circ \frac{dt}{1-t} 
\circ \cdots \circ 
\underbrace{\frac{dt}{t} \circ \frac{dt}{t} \circ \cdots \circ 
\frac{dt}{t}}_{k_n-1 \mbox{ times}} \circ \frac{dt}{1-t}
\;,  
\label{iterated}
\end{eqnarray}
where, by definition
\begin{eqnarray}
\int_0^z 
\underbrace{\frac{dt}{t} \circ \frac{dt}{t} \circ \cdots \circ 
\frac{dt}{t}}_{k_1-1 \mbox{ times}} \circ \frac{dt}{1-t} 
= 
\int_0^z 
\frac{dt_1}{t_1} \int_0^{t_1} \frac{dt_2}{t_2} \cdots 
\int_0^{t_{k-2}} \frac{dt_{k_1-1}}{t_{k_1-1}}
\int_0^{t_{k_1-1}} \frac{dt_{k_1}}{1-t_{k_1}} \;.
\end{eqnarray}
The integral (\ref{iterated}) is an iterated Chen integral \cite{Chen} 
w.r.t.\ the differential forms $\omega_0 = dz/z$ and 
$\omega_1 = \frac{dz}{1-z}$, so that 
\begin{eqnarray}
\Li{k_1, \cdots, k_n}{z} & = & \int_0^z \omega_0^{k_1-1} \omega_1 
\cdots \omega_0^{k_n-1} \omega_1 \;.
\label{chen}
\end{eqnarray}

\section{All-order $\ep$-expansion of generalized hypergeometric functions 
with integer values of parameters}
\label{allorder}

In this section, we shall prove {\bf Theorem 1}. We begin by noting that
Eq.~(\ref{reduction}) can be written in a slightly different form:
in terms of any basic function  ${}_{p}F_{p-1}(\vec{a};\vec{b}; z)$ and
its first $p-1$ derivatives,
\begin{eqnarray}
&& \hspace{-5mm}
R_{p+1}(\vec{a},\vec{b},z)\;
{}_{p}F_{p-1}(\vec{a}+\vec{m};\vec{b}+\vec{k}; z)
 = 
\nonumber \\ && \hspace{-5mm}
\Biggl \{
  R_1(\vec{a},\vec{b},z)\,\theta^{\,p-1}
+ \cdots 
+ R_{p-1} (\vec{a},\vec{b},z)\,\theta
+ R_{p}(\vec{a},\vec{b},z) 
\Biggr\}
{}_{p}F_{p-1}(\vec{a};\vec{b}; z) \;,
\label{decomposition:integer}
\end{eqnarray}
where 
$\vec{m}, \vec{k}$ are lists of integers, 
the $R_i$ are polynomials in parameters $a_i,b_j$ and $z$, 
and $\theta = z \frac{d}{d z} \;$. The essential step in proving 
{\bf Theorem 1} is the following lemma:\\

\noindent{\bf Lemma 1} \\
{\it 
The all-order $\ep$-expansion of the function 
${}_pF_{p-1}(\vec{a}\ep;\vec{1}+\vec{b}\ep;z)$, 
is expressible in terms of generalized polylogarithms $($Eq.\ \ref{GP}$)$.
} \\[1em]

Lemma 1 could be proved in the same manner as in case of multiple $($inverse$)$ 
binomial sums.  This was done in Ref.\ \cite{nested1}. 
However, it is fruitful to prove it using the construction of an iterated 
solution of the proper differential equation related to the hypergeometric 
function.\footnote{The 
proper solution for Gauss hypergeometric functions was constructed in 
Refs.\ \cite{hyper:expansion,KWY07:1}.}
We will follow this technique here, and in the process construct an 
iterative algorithm determining the analytical coefficients of the 
epsilon expansion.

Let us consider the differential equation for the hypergeometric function 
$\omega(z) = {}_pF_{p-1}(\vec{a}\ep;\vec{1}+\vec{b}\ep;z)$:
\begin{equation}
\left[ 
z \prod_{i=1}^p (\theta + a_i \ep)
- \theta \prod_{i=1}^{p-1} (\theta + b_i \ep)
\right] 
\omega(z) = 0 \;.
\label{diff}
\end{equation}
The boundary conditions for basis functions are 
$\omega(0)=1$ and $\left. \theta^j 
\omega(z)\right|_{z=0} = 0$, where $j=1, \cdots, p-1$.  The proper 
differential equation for $\omega(z)$ is valid in each order of $\ep$. 
Defining the coefficients functions $w_k(z)$ at each order by  
\begin{equation}
\omega(z) = \sum_{k=0}^\infty w_k(z) \ep^k, 
\label{epsilon-expansion}
\end{equation}
the boundary conditions for the coefficient functions are
\begin{subequations}
\label{w-bdry-conds}
\begin{eqnarray}
 & w_0(z) &= 1\;,  
\label{w0=1} \\
 & w_k(z) &= 0\;, \qquad k<0 \;,
\label{wneg} \\
& w_k(0) &= 0\;, \qquad  k \geq 1 \;, 
\label{w(z=0)} \\
& \left. z \frac{d}{dz} w_k(z) \right|_{z=0} &= 0\;, \qquad  k \geq 0 \; ,
\label{1:boundary} \\
& \cdots & 
\nonumber \\
& \left. \left( z \frac{d}{dz} \right)^{p-1} w_k(z) \right|_{z=0} &= 0\;, 
\qquad  k \geq 0 \; .
\label{integer:boundary}
\end{eqnarray}
\end{subequations}
The differential equation (\ref{diff}) has the form 
\begin{equation}
\left[(1\!-\!z) \frac{d}{dz} \right] \left( z \frac{d}{dz} \right)^{p-1} w_k(z)
= \sum_{i=1}^{p-1}
\left[ P_i(\vec{a}) \!-\! \frac{1}{z} Q_i(\vec{b}) \right] 
\left( z \frac{d}{dz} \right)^{p-i} w_{k-i}(z) + P_p(\vec{a}) w_{k-p}(z) \;, 
\nonumber \\ 
\label{general:diff}
\end{equation}
where $P_j(\vec{a})$ and $Q_j(\vec{b})$ are polynomials of order $j$
defined on spaces of $p$- and $(p\!-\!1)$-vectors $\vec{a}$ and 
$\vec{b}$, respectively.  They are defined as
\begin{subequations}
\begin{eqnarray}
P_0 & = & Q_0 = 1 \;, 
\\ 
P_r & = & \Sigma_{i_1,\cdots,i_r=1}^p 
\Pi_{i_1 < \cdots < i_r } a_{i_1} \cdots a_{i_r} \;, \quad r=1, \cdots, p\;, 
\\ 
Q_r & = & \Sigma_{i_1,\cdots,i_r=1}^{p-1} \Pi_{i_1 < \cdots < i_r } b_{i_1} 
\cdots  b_{i_r} \;, \quad r=1, \cdots, p-1\;, 
\\ 
Q_p & = & 0 \;, \quad
\label{PQ}
\end{eqnarray}
\end{subequations}
so that 
\begin{eqnarray}
P_1  =  \Sigma_{j=1}^p a_j \;,& \quad 
&Q_1 = \Sigma_{j=1}^{p-1} b_j \;, 
\nonumber \\ 
P_2  =  \Sigma_{i,j=1; i<j}^p a_i a_j \;,& \quad  
&Q_2 = \Sigma_{i,j=1; i < j}^{p-1} b_i b_j \;, 
\nonumber \\ 
P_3  =  \Sigma_{i_1,i_2,i_3=1; i_1 < i_2 < i_3 }^p a_{i_1} a_{i_2} a_{i_3}\;, 
& \quad
&Q_3 = \Sigma_{i_1,i_2,i_3=1; i_1 < i_2 < i_3 }^{p-1} b_{i_1} b_{i_2} b_{i_3}\;.
\nonumber \\ 
\cdots& \quad &\cdots
\\ 
&\qquad &Q_{p-1}  =  \Pi_{i=1}^{p-1} b_i \;, \quad 
\nonumber \\ 
 P_p  =  \Pi_{i=1}^p a_i \;,& \qquad
& Q_p = 0.
\nonumber 
\end{eqnarray}
The polynomials $P_j$ and $Q_j$ satisfy the following relations: 
\begin{eqnarray}
P_j(\vec{a},b) & = & P_j(\vec{a}) + b P_{j-1} (\vec{a}) \;, \quad 
Q_j(\vec{a},b)  =  Q_j(\vec{a}) + b Q_{j-1} (\vec{a}), \quad j=1, \cdots, p \;.
\end{eqnarray}
In particular, 
\[
P_j(\vec{a},0) =  P_j(\vec{a}) \;, \quad 
Q_j(\vec{a},0)  =  Q_j(\vec{a})  \;.
\]

Let us introduce a set of a new functions $\rho^{(j)}(z), j=1, \cdots,p-1$ 
defined by
\begin{eqnarray}
\rho^{(j)}(z)  & = &  
\theta^j \omega(z) 
\equiv \left( z \frac{d}{dz} \right)^j \omega(z) 
= \sum_{k=0}^\infty \rho^{(j)}_k(z)  \ep^k
\;, \quad j=1, \cdots, p-1 \;,
\end{eqnarray}
where the coefficient functions $\rho^{(j)}_k(z)$ satisfy 
\begin{equation}
\rho^{(j)}_k(z) =  \left( z \frac{d}{dz} \right)^j 
 w_k(z) \;, \quad j=1, \cdots, p-1 \;.
\end{equation}
The boundary conditions for these new functions follow from 
Eq.\ (\ref{integer:boundary}):
\begin{equation}
\rho^{(j)}_k(0)  =  0 \;, \qquad k \geq 0 \;, \quad j \geq 1 \;.
\label{integer:boundary:rho}
\end{equation}
Eq.\ (\ref{general:diff}) can be rewritten as a system of first-order 
differential equations
\begin{subequations}
\label{integer:diff}
\begin{eqnarray}
z \frac{d}{dz} \rho^{(j)}_k(z) & = & \rho^{(j+1)}_k(z) \;, \quad j=0, 1, 
\cdots, p-1
\label{integer:diff1}
\\ 
(1-z) \frac{d}{dz} \rho^{(p-1)}_k (z) & = &  
\sum_{i=1}^p \left[ P_i(\vec{a}) \!-\! \frac{1}{z} Q_i(\vec{b}) \right] 
\rho^{(p-i)}_{k-i}(z) \;,
\label{integer:diff2}
\end{eqnarray}
\end{subequations}
and we have
\begin{equation}
w_k(z) \equiv  \rho^{(0)}_k(z) \;.
\label{w0}
\end{equation}
The solution of system (\ref{integer:diff}) can be presented in an iterated 
form: 
\begin{subequations}
\begin{eqnarray}
\rho^{(p-1)}_k (z)
& = &  
\sum_{i=1}^p
\left[ P_i(\vec{a}) \!-\! Q_i(\vec{b}) \right] \int_0^z \frac{dt}{1-t} 
\rho^{(p-i)}_{k-i}(t)
\nonumber \\ 
& - & 
\sum_{i=1}^{p-2} Q_i(\vec{b}) \rho^{(p-i-1)}_{k-i}(z) 
- Q_{p-1}(\vec{b}) [w_{k-p+1}(z)  \!-\! \delta_{0,k-p+1}] \;,
\label{integer:w:a}
\\ 
\rho^{(j-1)}_k (z) & = & \int_0^z \frac{dt}{t} \rho^{(j)}_k(t) \;, \quad k 
\geq 1 \;, \quad j=1, 2, \cdots, p \!-\! 1 \;,
\label{integer:w:b}
\end{eqnarray}
\label{integer:w}
\end{subequations}
where $\delta_{a,b}$ is the Kronecker delta function.

From the system of Eq.~({\ref{integer:w}}), it is easy to find that 
\begin{equation}
\rho_k^{(j)}(z) = 0, \quad k < p; \quad j=0,1,\cdots,p-1.
\label{zero}
\end{equation}
The first nonzero terms are generated by Eq.~(\ref{integer:w:a}) for $i=k=p$.
Substituting this result into Eq.~(\ref{integer:w:b}) we will find the solution 
of the first iteration:
\begin{equation}
\rho_p^{(p-1-j)}(z) = P_p(\vec{a}) \Li{1+j}{z} \;, \quad j=0,1,\cdots,p-1, 
\label{iteration:first}
\end{equation}
where $\Li{j}{z}$ is a classical polylogarithm\,\cite{Lewin}  and
$
\Li{1}{z} = - \ln(1-z).
$
{\bf Lemma 1} follows from the representation (\ref{integer:diff2}), the 
value $w_0(z)=1$, the definition of generalized polylogarithms (\ref{GP}),
and  Eq.~(\ref{iteration:first}).

The case when one of the upper parameters of the hypergeometric function
is a positive integer number,
${}_pF_{p-1}(I_1, \vec{A}+\vec{a}\ep;\vec{B}+\vec{b}\ep;z)$, 
corresponds to $a_1$ equal to zero. 
A smooth limit exists in this case and the particular result
can be reproduced from expression (\ref{integer:w}). 
{\bf Theorem 1} is thus proved.

\section{Explicit expressions for the first five coefficients of the expansion}
Let us return to Eq.~({\ref{integer:w}}) and look at the next terms of the
expansion.  The first nonzero term of the iteration is given by 
Eq.~(\ref{zero}).  The second iteration corresponds to $k=p+1$. 
In the r.h.s.\ of Eq.~(\ref{integer:w:a}), only terms with $i=1$ produce 
a non-zero contribution,
\[
\frac{\rho_{p+1}^{(p-1)}(z)}{P_p} = 
\Delta_1  \frac{1}{2} \ln^2 (1-z) - Q_1 \Li{2}{z} \;,
\]
where for simplicity, we omit arguments in the functions $P_j,Q_j$ and 
introduce a notation
\[
\Delta_j = P_j - Q_j \;, \quad j=1,\cdots,p-1.
\]
Substituting the results in Eq.~(\ref{integer:w:b}),
we will get the solution of the second iteration:
\begin{eqnarray}
\frac{\rho_{p+1}^{(p-1-j)}(z)}{P_p} & = & 
\Delta_1 \Li{j+1,1}{z} 
\!-\! 
Q_1 \Li{2+j}{z} 
\;, \quad j=0,1,\cdots,p-1,
\label{iteration:second:b}
\end{eqnarray}
where $\Li{a_1,\cdots, a_k}{z}$ is a generalized polylogarithm.  
The third iteration corresponds to $k=p+2$, and 
in the r.h.s. of Eq.~(\ref{integer:w:a}) only terms with $i=1,2$ will 
produce a non-zero contribution,  
\begin{eqnarray}
\frac{\rho_{p+2}^{(p-1-j)}(z)}{ P_p} & = & 
\Delta_1^2 \Li{j+1,1,1}{z} 
\!+\! 
\left( \Delta_2 - Q_1 \Delta_1 \right) \Li{j+1,2}{z} 
\nonumber \\ & +  & 
\left(Q_1^2 \!-\!  Q_2 \right)  \Li{j+3}{z} 
\!-\! Q_1 \Delta_1  \Li{j+2,1}{z} \;, \quad j=0,1,\cdots,p-1.
\label{third}
\end{eqnarray}
The fourth iteration corresponds to $k=p+3$ and equal to 
\begin{eqnarray}
\frac{\rho_{p+3}^{(p-1-j)}(z)}{ P_p} & = & 
\Delta_1^3 \Li{j+1,1,1,1}{z} 
+ 
\Delta_1 
\left( \Delta_2 - Q_1 \Delta_1 \right)
\left[ 
\Li{j+1,1,2}{z} 
+ 
\Li{j+1,2,1}{z} 
\right]
\nonumber \\ && 
+ 
\left( \Delta_1 Q_1^2 - \Delta_1 Q_2 - \Delta_2 Q_1 + \Delta_3 \right)
\Li{j+1,3}{z} 
- Q_1 \Delta_1^2 \Li{j+2,1,1}{z} 
\nonumber \\ && 
+ Q_1 \left( \Delta_1 Q_1 - \Delta_2 \right) \Li{j+2,2}{z} 
+ \Delta_1 \left(Q_1^2 - Q_2\right) \Li{j+3,1}{z} 
\nonumber \\ && 
- 
\left(Q_1^3 - 2Q_1 Q_2 + Q_3\right) \Li{j+4}{z} \;, 
\quad j=0,1,\cdots, p-1.
\label{fourth}
\end{eqnarray}
The fifth iteration corresponds to $k=p+4$ and equal to 
\begin{eqnarray}
\frac{\rho_{p+4}^{(p-1-j)}(z)}{ P_p} & = & 
\Delta_1^4 \Li{j+1,1,1,1,1}{z} 
+
\left( \Delta_1^2 Q_1^2 - 2 \Delta_1 \Delta_2 Q_1 + \Delta_2^2 \right)
\Li{j+1,2,2}{z} 
\nonumber \\ && 
+ 
\Delta_1^2 
\left( \Delta_2 - Q_1 \Delta_1 \right)
\left[ 
\Li{j+1,1,1,2}{z} 
+ 
\Li{j+1,1,2,1}{z} 
+ 
\Li{j+1,2,1,1}{z} 
\right]
\nonumber \\ && 
+
\Delta_1 
\left\{ \Delta_1 \left(Q_1^2 - Q_2\right) - \Delta_2 Q_1 + \Delta_3 \right\}
\left[ 
\Li{j+1,1,3}{z} 
+
\Li{j+1,3,1}{z} 
\right]
\nonumber \\ && 
- Q_1 \Delta_1^3 \Li{j+2,1,1,1}{z} 
+ 
Q_1 \Delta_1 
\left( \Delta_1  Q_1 - \Delta_2 \right)
\left[ 
\Li{j+2,1,2}{z} 
+ 
\Li{j+2,2,1}{z} 
\right]
\nonumber \\ && 
+ \Delta_1^2 \left(Q_1^2 - Q_2\right) \Li{j+3,1,1}{z} 
- \Delta_1 
\left( Q_1^3 - 2 Q_1 Q_2 + Q_3 \right) \Li{j+4,1}{z} 
\nonumber \\ && 
+ 
Q_1
\left\{  \Delta_1 \left(Q_2 - Q_1^2\right) + \Delta_2 Q_1 - \Delta_3  \right\} \Li{j+2,3}{z} 
\nonumber \\ && 
+ 
\left[ Q_1 \left\{  \Delta_1 \left(Q_2 - Q_1^2\right) + \Delta_2 Q_1 \right\} - Q_2 \Delta_2 \right]
\Li{j+3,2}{z} 
\nonumber \\ && 
+ 
\left( Q_1^4 - 3 Q_1^2 Q_2 + 2Q_1 Q_3 + Q_2^2 - Q_4 \right)
\Li{j+5}{z} 
\nonumber \\ && 
+
\left\{
\Delta_4 - Q_1 \Delta_3  + \Delta_2 \left(Q_1^2 - Q_2\right) 
-  \Delta_1  \left(Q_1^3 - 2 Q_1 Q_2 + Q_3\right)
\right\}
\Li{j+1,4}{z} ,
\nonumber\\
& & \quad j=0,1,\cdots, p-1.
\label{fifth}
\end{eqnarray}
For lower values of the index $p$, the following relations can be used for 
transforming harmonic polylogarithms\,\cite{RV00} to the 
classical\,\cite{Lewin} or Nielsen\,\cite{Nielsen} ones: 
\begin{subequations}
\begin{eqnarray}
&& 
\Li{j,\underbrace{1,1,\cdots,1}_{p \mbox{ times}}}{z} = \Snp{j-1,p+1}{z} \;,
\\ && 
\Snp{0,j}{z} =  \frac{(-1)^j}{j!} \ln^j(1-z) \;,
\\ && 
\Li{1,2}{z}  =  - \ln(1-z) \Li{2}{z} - 2 \Snp{1,2}{z} \;,
\\ && 
\Li{1,3}{z}  =  -\ln(1-z) \Li{3}{z} - \frac{1}{2} \left[ \Li{2}{z} \right]^2 \;,
\\ && 
\Li{1,4}{z}  =  -\ln(1-z) \Li{4}{z} + F_2(z) \;,
\\ && 
\Li{2,2}{z}  =  \frac{1}{2} \left[ \Li{2}{z} \right]^2 - 2 \Snp{2,2}{z} \;,
\\ && 
\Li{3,2}{z}  =  \frac{1}{2} \Li{2}{z} \Li{3}{z} + \frac{1}{2} F_2(z) 
- 2 \Snp{3,2}{z} \;,
\\ && 
\Li{2,3}{z}  =  - \frac{3}{2} F_2(z) - \frac{1}{2} \Li{2}{z} \Li{3}{z} \;.
\\ && 
\Li{1,1,2}{z}  =  \frac{1}{2} \ln^2(1-z)  \Li{2}{z} + 2 \ln(1-z) \Snp{1,2}{z} 
+ 3 \Snp{1,3}{z} \;,
\\ && 
\Li{1,2,1}{z}  =  - \ln(1-z) \Snp{1,2}{z} - 3 \Snp{1,3}{z} \;, 
\\ && 
\Li{2,2,1}{z} + \Li{2,1,2}{z} =  F_1(z) - \Li{2}{z} \Snp{1,2}{z} \;, 
\\ && 
\Li{1,1,3}{z} + \Li{1,3,1}{z} = 
\frac{1}{2} \ln^2 (1-z) \Li{3}{z} 
+ \frac{1}{2} \ln (1-z) \left[ \Li{2}{z} \right]^2
\nonumber \\ && \hspace{30mm}
- \ln(1-z) \Snp{2,2}{z} 
- \Li{2}{z} \Snp{1,2}{z} 
+ F_1(z) \;,
\\ && 
\Li{1,2,2}{z}  = 
\ln(1-z) \left\{ 2 \Snp{2,2}{z} - \frac{1}{2} \left[ \Li{2}{z} \right]^2 
\right\} 
\nonumber \\ && \hspace{30mm}
+ 2 \Li{2}{z} \Snp{1,2}{z} - 2 F_1(z) \;, 
\\ && 
\Li{1,1,1,2}{z} + \Li{1,1,2,1}{z} + \Li{1,2,1,1}{z} = 
- \frac{1}{6} \ln^3 (1-z) \Li{2}{z} 
\nonumber \\ && \hspace{30mm}
- \frac{1}{2} \ln^2 (1-z) \Snp{1,2}{z} 
- \ln (1-z) \Snp{1,3}{z} 
- 2 \Snp{1,4}{z} \;,
\end{eqnarray}
\end{subequations}
where we have introduced two new functions related algebraically 
(see Eqs.\ (2.23) -- (2.25) in Ref.\ \cite{KWY07:1}):
\begin{eqnarray}
F_1(z) & = & \int_0^z \frac{dx}{x} \ln^2(1 - x) \Li{2}{x} \; , 
\\
F_2(z) & = & \int_0^z \frac{dx}{x} \ln(1 - x) \Li{3}{x} \; .
\end{eqnarray}

For completeness, we will present the values of $P$ and $Q$ for $p=3,4$:
\begin{itemize}
\item
$p=3$
\begin{eqnarray}
P_1(\vec{a}) & = & a_1 + a_2 + a_3 \;, \hspace{15mm} Q_1(\vec{b})  
=  b_1 + b_2 \;,
\nonumber \\ 
P_2(\vec{a}) & = & a_1 a_2 + a_1 a_3 + a_2 a_3 \;, \quad Q_2(\vec{b})  
=  b_1 b_2 \;.
\nonumber \\ 
P_3(\vec{a}) & = & a_1 a_2 a_3 \;,  \hspace{25mm} Q_3(\vec{b}) =  0 \;.
\end{eqnarray}

\item
$p=4$
\begin{eqnarray}
P_1(\vec{a}) & = & a_1 \!+\! a_2 \!+\! a_3 \!+\! a_4 \;, 
\hspace{40mm} 
Q_1(\vec{b})  =  b_1 \!+\! b_2 \!+\! b_3\;,
\nonumber \\ 
P_2(\vec{a}) & = & a_1 a_2 \!+\! a_1 a_3 \!+\! a_1 a_4 \!+\! a_2 a_3 
\!+\! a_2 a_4 \!+\! a_3 a_4\;, 
\quad 
Q_2(\vec{b})  =  b_1 b_2 \!+\! b_1 b_3 \!+\! b_2 b_3 \;.
\nonumber \\ 
P_3(\vec{a}) & = & a_1 a_2 a_3 \!+\!  a_1 a_2 a_4 \!+\!  a_2 a_3 a_4 \;,  
\hspace{25mm} 
Q_3(\vec{b}) =  b_1 b_2 b_3 \;.
\nonumber \\ 
P_4(\vec{a}) & = & a_1 a_2 a_3 a_4 \;,  
\hspace{50mm} 
Q_4(\vec{b}) =  0 \;.
\end{eqnarray}
\end{itemize}
The first few coefficients, up to order 4, could be cross-checked using the 
results of Ref.\ \cite{FKV99}.

We would like to point out that Eqs.\ (\ref{iteration:second:b}) -- 
(\ref{fifth}) contain an explicit logarithmic singularity at $z=1$. It is 
well-known that the generalized hypergeometric function 
${}_pF_{p-1}(\vec{a};\vec{b};z)$ 
converges absolutely on the unit circle $|z|=1$ if 
\[
{\rm Re}\; \left( \sum_{j=1}^{p-1} b_j - \sum_{j=1}^p a_j \right) > 0.
\]
In this case, the coefficients of the $\ep$-expansion also converge at 
each order in $\ep$. To get a smooth limit, it is enough to rewrite 
Eqs.\ (\ref{iteration:second:b}) -- (\ref{fifth}) in terms of 
functions of argument $1-z$ and set $z=1$.

\section{Conclusions}

We have shown ({\bf Theorem 1}) that the $\ep$-expansions of 
generalized hypergeometric functions with integer values of parameters
are expressible in terms of generalized polylogarithms (see Eq.~(\ref{GP}))
with coefficients that are ratios of polynomials.
The proof includes (i) the differential reduction algorithm;
and (ii) iterative algorithms for calculating the analytical
coefficients of the $\ep$-expansion of basic hypergeometric functions 
(see Eq.~(\ref{integer:w})).  The first five coefficients of the 
$\ep$-expansion for basis hypergeometric functions
are calculated explicitly in Eqs.~(\ref{iteration:first}),
(\ref{iteration:second:b}), (\ref{third}), (\ref{fourth}), and (\ref{fifth}).  
The FORM\,\cite{FORM} representations of these expressions and the next 
coefficients are available via Ref.\ \cite{MKL}.
\acknowledgments 

This research was supported by NATO Grant PST.CLG.980342 and 
DOE grant DE-FG02-05ER41399. 
M.\ Yu.\ K. is supported in part by BMBF 05 HT6GUA, and is thankful to Baylor 
University for support of this research, and very grateful to his wife, 
Laura Dolchini, for moral support while working on the paper.

\newpage

\end{document}